\documentclass[10pt]{article}

\AtBeginDocument{%
  }

\usepackage[utf8]{inputenc}
\usepackage[T1]{fontenc}
\usepackage{lmodern}
\usepackage{newunicodechar}
\newunicodechar{̈}{"{}}
\usepackage{geometry}
\usepackage{setspace}
\usepackage{microtype}
\usepackage{hyperref}
\usepackage{graphicx}
\usepackage{authblk}
\usepackage{longtable}
\usepackage{booktabs}
\usepackage{array}
\usepackage{enumitem}
\usepackage{url}
\usepackage{ifthen}
\usepackage{xcolor}
\usepackage{times} 
\usepackage{tgbonum}
\usepackage{caption}

\hypersetup{colorlinks=true, linkcolor=black, urlcolor=black, citecolor=black}

\RequirePackage[
  datamodel=acmdatamodel,
  style=acmnumeric,
  sorting=nyt,
  maxbibnames=50,
  ]{biblatex}

\makeatletter
\renewcommand{\fnum@figure}{Figure \thefigure}

\makeatother

\newcounter{BHNumberOfComments}
\stepcounter{BHNumberOfComments}

\newcolumntype{P}[1]{>{\raggedright\arraybackslash}p{#1}}

\title{Why Today’s Internet Cannot Support Tomorrow’s Immersive Experiences: Insights from \\the NSF Workshop}

\author[1]{Bo Han}
\author[2]{Amitabh Varshney}
\author[3]{Klara Nahrstedt}
\author[1]{Songqing Chen}
\author[3]{Sarita Adve}
\author[4]{Ruofei Du}
\author[5]{Xueyu Hou}
\author[5]{Yongjie Guan}
\author[6]{Jiayi Meng}
\author[7]{Qiao Jin}
\author[1]{Xiaonan Guo}
\author[1]{Nan Wu}
\author[1]{Ruizhi Cheng}

\affil[ ]{%
\begin{tabular}{@{}l l l@{}}
$^1$George Mason University,
$^2$University of Maryland,
$^3$University of Illinois at Urbana-Champaign,
$^4$Google
$^5$University of Maine,
$^6$The University of Texas at Arlington,
$^7$North Carolina State University

\end{tabular}
}

\date{} 

\begin{document}
\maketitle

\vspace{-0.7cm}
\begin{center}
{\small\textbf{Corresponding Author:} Bo Han, \href{mailto:bohan@gmu.edu}{bohan@gmu.edu}.}
\end{center}

\begin{abstract}

Immersive computing is poised to become a major paradigm for future human--computer interaction (HCI), enabling spatial and embodied experiences across applications such as education, healthcare, manufacturing, emergency response, and scientific discovery. 
However, today's Internet and computing infrastructure were not designed for real-time, multi-user, sensor-rich, and privacy-sensitive immersive applications. 
They require coordination across sensing, representation, compression, networking, edge/cloud computing, rendering, security/privacy, and human experience. This report summarizes key findings from the {\em NSF Workshop on Networking and Systems Challenges in Immersive Computing}. 
The workshop brought together researchers from networking, systems, extended reality (XR), artificial intelligence (AI), graphics, HCI, security, privacy, and applications to identify challenges and opportunities. 
We conclude that immersive computing cannot be enabled by isolated advances in network bandwidth, device capabilities, AI models, or rendering alone. 
The field needs end-to-end architectures that adapt to user behavior, application requirements, resource constraints, and social expectations. 
We organize the agenda into six themes: networked systems for immersive computing; real-time edge--cloud orchestration for XR applications; AI for immersive experiences; privacy, security, and trust in XR systems; human-centered design and user experience in XR; and scalable infrastructure, testbeds, and open ecosystems for XR. We further identify key research questions, future trends, risks, ten-year success metrics, and action items for academia, industry, government, and research community. The report calls for coordinated investment in open platforms, reproducible benchmarks, privacy-preserving systems, and human-centered evaluation to build trustworthy, scalable, and broadly impactful infrastructure.

\end{abstract}

\maketitle

\section*{Executive Summary}
Immersive computing~\cite{han2019mobile} is emerging as a new paradigm for how people interact with digital information, physical environments, remote collaborators, and artificial intelligence (AI) agents. Unlike traditional Internet applications that primarily deliver text, audio, images, or video to users, immersive systems must continuously sense the world~\cite{wu2025pipe}, interpret user embodied behavior~\cite{han2020vivo}, transmit spatial and semantic information~\cite{cheng2024magicstream}, and adapt to changing human and network conditions in real time~\cite{zhang2022m5}. 
However, today's Internet and computing stack were not originally built for immersive experiences. They were designed mainly for screen-based interaction and device-centered computing, while immersive systems require real-time multi-user coordination, continuous sensing of people and environments, spatially grounded interaction, and stronger protections for sensitive social and behavioral data.

The {\em NSF Workshop on Networking and Systems Challenges in Immersive Computing} brought together researchers across networking, systems, extended reality (XR), AI, graphics, human--computer interaction (HCI), security, privacy, and application domains to identify the key barriers and opportunities for the next decade. 
The workshop converged on the following major insights and recommendations.

\begin{enumerate}[leftmargin=*]

\item \textbf{Rethinking the Internet for immersive experiences.}
Today's Internet was designed to move bits quickly, not to support the real-time, context-sensitive demands of immersive interaction.
Future immersive networking must go beyond throughput to help systems determine what to transmit, where to compute, and how to prioritize content based on user attention and application context~\cite{cheng2024magicstream,han2020vivo,lee2023farfetchfusion,lee2020groot,wu2024theia}.
This requires a shift toward context-aware, application-aware, and privacy-aware network architectures~\cite{cheng2024metafl,Stephenson2022sok,wu2025pipe}.

\item \textbf{Moving from raw data streaming to semantic and adaptive communication.}
Current immersive systems often capture and transmit substantially more visual and sensor data than individual tasks actually require, leading to unnecessary bandwidth consumption, energy overhead, and privacy exposure.
Future systems should transmit task-relevant information, such as scene structure, user intent, object semantics, or prioritized content, rather than raw streams~\cite{cheng2024magicstream,yang2025empilo}.
The long-term goal is not to send more pixels, but to send the right information at the right time.

\item \textbf{Designing real-time edge--cloud orchestration for XR.}
Unlike video streaming, immersive interaction is latency-sensitive and leaves little room for buffering, making compute placement a first-order systems problem.
Immersive systems must dynamically distribute work across smartglasses, phones, edge devices, and cloud servers based on latency, power, privacy, and application needs~\cite{kong2023accumo}.
The field currently lacks principled models and standards for XR offloading, and closing this gap is essential for practical deployment.

\item \textbf{Building AI-native immersive systems carefully.}
AI offers powerful capabilities for perception, compression, personalization, and intelligent interaction in XR~\cite{han2026ai}, but it also introduces computational latency, energy costs, and generalization risks.
Machine learning is not always superior to geometric or rule-based methods, particularly in time-critical tasks where predictability matters.
Future systems should use AI selectively and interpretably, co-designed with networking, rendering, and human-centered constraints rather than added as an afterthought.

\item \textbf{Making security, privacy, and trust first-class design goals.}
XR devices continuously sense the wearer and the surrounding environment through cameras~\cite{wu2025pipe}, microphones~\cite{Anderson2026sound}, eye-tracking~\cite{corbett2023bystandar}, hand-tracking~\cite{cheng2024metafl}, and spatial mapping~\cite{wu2022privacy}, thereby creating privacy risks that far exceed those of conventional mobile devices.
These capabilities expose users and bystanders alike to identity leakage and behavioral inference.
Security and privacy must therefore be built into the full system stack from the start, not patched in at the application layer.

\item \textbf{Centering human experience and social acceptability.}
Technical performance alone will not determine whether immersive computing succeeds; devices must be comfortable, inclusive, and socially acceptable in real-world settings.
The workshop emphasized the need for longitudinal in-situ studies~\cite{jin2026we}, broad stakeholder involvement~\cite{jin2022will}, and metrics that capture cognitive load, comfort, accessibility, trust, and social norms~\cite{Creed2024Inclusive,krokos2022quantifying}, not just frame rate or bandwidth.
Human-centered evaluation must become a core research practice, not an optional add-on.

\item \textbf{Creating open testbeds, benchmarks, and reproducible ecosystems.}
Immersive systems research is currently hindered by the lack of shared infrastructure that reflects real users, heterogeneous devices, and realistic network conditions.
Open platforms such as CoMIC~\cite{han2023comic} and ILLIXR~\cite{illixr}, together with XR benchmarking efforts~\cite{Popescu2025Benchmarking}, provide an important foundation, but significant gaps remain.
Sustained community investment in open testbeds, datasets, and reproducible benchmarks is essential for the field to move beyond isolated prototypes.

\end{enumerate}

\section{Background and Motivation}

\subsection{The Rise of Immersive Computing}

Immersive computing is evolving from a collection of specialized demonstrations into a broader computing paradigm in which users interact with digital information, remote collaborators, physical environments, and AI agents through spatial, embodied, and context-aware interfaces. 
Unlike traditional multimedia systems that primarily deliver two-dimensional audiovisual content to passive viewers, emerging immersive systems aim to support interactive, real-time, and personalized experiences across virtual reality (VR), augmented reality (AR), and mixed reality (MR), collectively referred to as extended reality (XR). 
This transition reframes immersive computing as both a next-stage Internet~\cite{cheng2022will} and a systems challenge: the goal is no longer simply to stream content to users, but to continuously sense, understand, compute, transmit, render, and adapt shared environments in response to user behavior and application context. 
More fundamentally, immersive computing represents a paradigm shift from content consumption to interactive environments enabled by volumetric video~\cite{han2020vivo,lee2020groot,wu2024theia}, neural representations~\cite{chen2024nerfhub,wu2025nevo}, telepresence~\cite{cheng2024magicstream,guan2023metastream,lee2023farfetchfusion}, and generative AI interaction~\cite{cheng2025hello}.

Motivated by this emerging transition, the {\em NSF Workshop on Networking and Systems Challenges in Immersive Computing} brought together researchers and practitioners from networking, systems, XR, AI, graphics, human--computer interaction (HCI), security, privacy, and application domains to examine the foundational challenges that must be addressed for immersive computing to become scalable, trustworthy, and broadly deployable. 
Discussions throughout the workshop highlighted how immersive computing can transform applications requiring rich spatial understanding, interactive learning, and real-time collaboration. In medical education, immersive environments can convert rare or difficult-to-access training scenarios into portable, repeatable, and scalable learning experiences. Prior studies have shown that VR-based surgical training can achieve learning outcomes comparable to cadaver-based instruction while improving accessibility, personalization, and opportunities for repeated practice~\cite{groves2025assessing}. These examples illustrate how immersive systems can bridge the gap between textbook knowledge and hands-on experience through spatial visualization, multimodal guidance, and interactive skill development.

At the same time, these applications expose the scale of the underlying systems challenge. Modern immersive capture platforms such as the University of Maryland HoloCamera studio~\cite{heagerty2024holocamera} use hundreds of cameras to generate high-fidelity 3D content, producing more than 100 billion pixels per second of data. Processing even short segments of such content can require many GPU-hours, creating substantial challenges for acquisition, representation, analysis, compression, storage, streaming, and real-time rendering. These observations reinforce a central theme of the workshop: immersive computing bottlenecks extend far beyond networking or display hardware, spanning the entire end-to-end pipeline, including sensing, computation, communication, and content representation.

The vision of all-day smartglasses further amplifies these challenges. Workshop discussions highlighted the possibility that lightweight AR glasses could eventually become a mainstream computing platform comparable in impact to smartphones, while also exposing major unresolved barriers related to form factor, wide-field-of-view displays, privacy-preserving tracking, content generation, 3D model inference, communication, and all-day battery life. Addressing these constraints will likely require computation to be dynamically partitioned across wearable devices, nearby companion hardware, edge infrastructure, and cloud resources to balance latency, energy consumption, privacy, and usability. Realizing this vision, therefore, requires not only advances in XR hardware but also new system architectures for wearable, context-aware, and continuously networked immersive computing.

Taken together, these examples demonstrate that immersive computing is not merely an entertainment technology. It has the potential to become foundational infrastructure for education, training, healthcare, collaboration, accessibility, and human-centered digital interaction.

\begin{figure}
    \centering
    \includegraphics[width=1.0\linewidth]{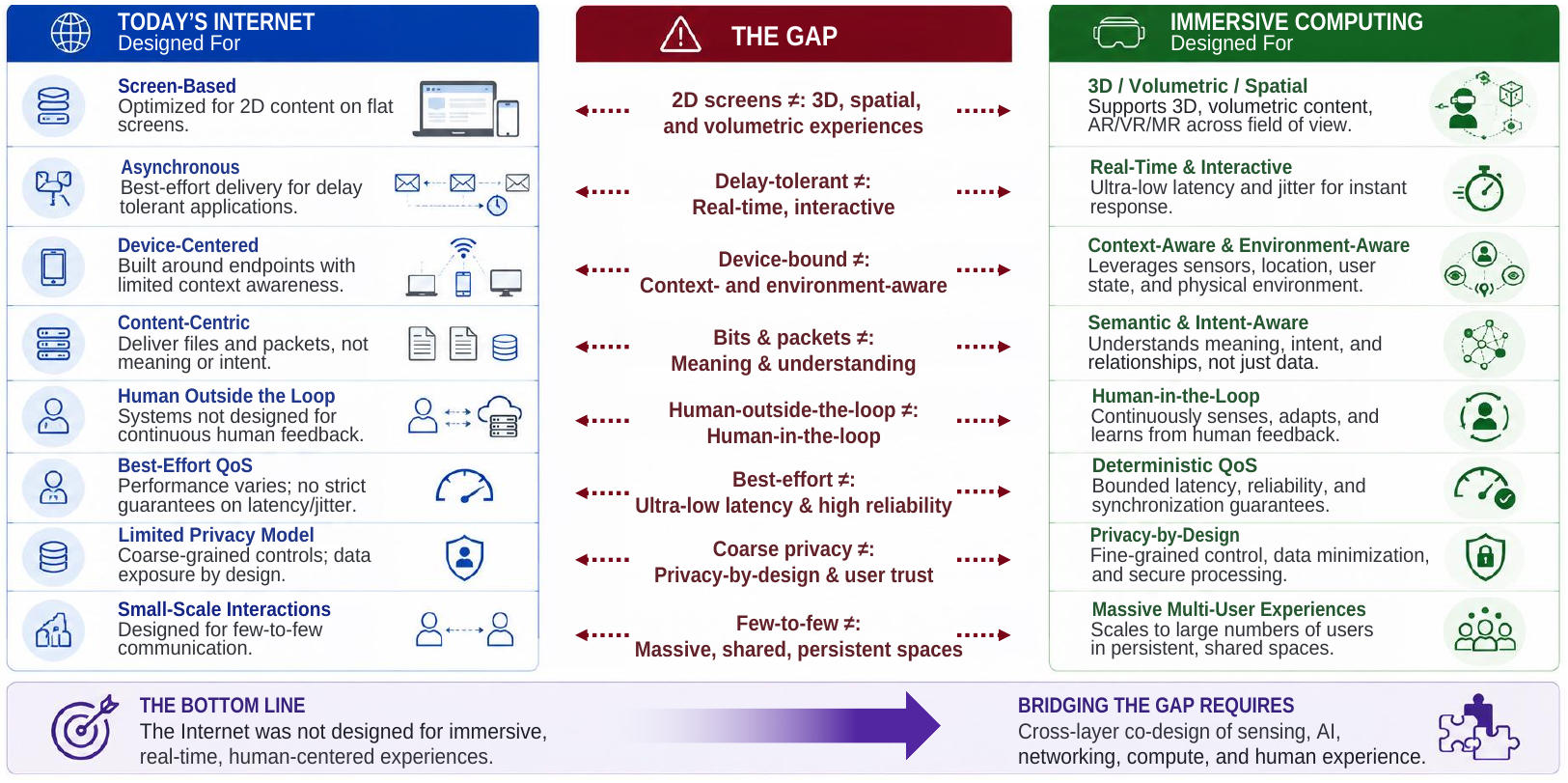}
    \caption{The Immersive Computing Gap: Why today's Internet falls short of tomorrow's immersive experiences.
    }
    \label{fig:gap}
\end{figure}

\subsection{Why Today’s Internet Falls Short}
Despite rapid progress in devices, graphics, AI, and wireless networks, today’s Internet and computing stack were not designed to meet the demands of real-time, immersive interaction. Conventional networked applications largely tolerate temporary stalls, delayed synchronization, or asynchronous interaction. 
However, immersive applications require tight coordination among sensing, tracking, computation, communication, rendering, and human perception. A missed frame, an inaccurate pose estimate, a delayed haptic cue, a privacy leak, or an inconsistent shared state can directly degrade presence, comfort, trust, and safety.
Figure~\ref{fig:gap} highlights the growing mismatch between the design assumptions of traditional Internet systems and the emerging requirements of immersive computing environments.

The workshop discussions emphasized that networking for immersive computing cannot be reduced to faster bit delivery alone. Instead, immersive systems require networks that determine what information should be transmitted, where computation and rendering should occur, and which resources are trustworthy and capable of supporting real-time interaction. XR devices continuously generate and consume diverse data streams, including user poses, spatial maps, sensor inputs, rendering outputs, semantic scene representations, and application state. These requirements motivate a broader architectural shift from traditional address-centric communication toward data-, context-, and trust-aware networking that can dynamically discover, prioritize, and coordinate distributed resources and information flows.

The edge/cloud computing breakout session reached a similar conclusion. Participants noted that immersive applications differ fundamentally from traditional video streaming because immersive experiences depend on real-time responsiveness rather than passive content delivery. The group identified unresolved questions about where computation should occur, on glasses, phones, nearby edge devices, base stations, cloudlet edge servers, or remote cloud infrastructure, and observed that a definitive architectural equilibrium has not yet materialized across performance, latency, energy consumption, storage overhead, and system efficiency. The breakout session also highlighted the lack of standards for XR offloading, the absence of principled quality-of-experience (QoE) models, and the need for dynamic, application-specific edge architectures.

Immersive computing also changes the meaning of data efficiency. Traditional systems often capture or transmit raw visual data and optimize later in the pipeline. In XR, this approach is increasingly untenable. The networked systems breakout session asked what actually needs to be captured, transmitted, processed, or rendered, noting that today’s systems often capture everything with cameras, even though much visual data is discarded during processing. Participants discussed event cameras, edge-detecting sensors, Wi-Fi, Bluetooth, ultra-wideband (UWB), mmWave, and hybrid sensing as possible ways to reduce data movement and improve privacy.

AI introduces both opportunities and new bottlenecks. Modern immersive systems increasingly rely on machine learning for perception, reconstruction, compression, prediction, evaluation, personalization, and agentic interaction. However, the AI breakout session emphasized that AI is not automatically better: many models are optimized for quality rather than latency, can add computational and transmission overhead, and may generalize poorly across changing lighting, motion, environments, and users. In time-critical XR applications, traditional geometric methods may still outperform machine learning methods unless AI is carefully co-designed with system and network constraints.
Thus, AI-native immersive computing~\cite{han2026ai} requires not only better models, but also new abstractions for deciding when to use AI, where to run it, what data to expose to it, and how to evaluate its effect on human experience.

Security and privacy further expose the limitations of today’s computing and networking architectures. XR devices continuously sense both users and surrounding environments through gaze tracking, motion, voice, gestures, biometrics, spatial mapping, and multimodal interaction, while also capturing nearby people, objects, and activities. Workshop discussions highlighted that bystander visual data can reveal sensitive information and that immersive systems may require mechanisms to distinguish intended interaction partners from uninvolved bystanders in real time. At the same time, the sensing capabilities that create new privacy risks may also enable privacy-aware interaction, selective filtering, and context-aware access control. These concerns are not peripheral; they are likely to play a central role in determining whether immersive computing can achieve broad social acceptance and widespread real-world deployment.

\subsection{Fragmentation in Current Research}

A central motivation for the workshop was the recognition that immersive computing spans multiple research communities that often operate in isolation. Networking and systems researchers focus on latency, wireless communication, resource allocation, runtime partitioning, and edge/cloud orchestration; graphics, vision, and AI researchers study capture, rendering, reconstruction, multimodal models, and generative representations; while HCI, security, and privacy researchers examine human experience, accessibility, trust, safety, and sensing risks. Yet immersive systems require all of these concerns to be addressed jointly rather than independently.

This fragmentation is reflected in the lack of shared abstractions, evaluation methodologies, and cross-layer design principles. Workshop discussions highlighted that XR QoE models remain underdeveloped, many systems still rely heavily on heuristics, and existing AI models often fail to capture the nuances of immersive interaction beyond static visual understanding. Participants also emphasized that core problems such as sensing, naming, discovery, transmission, compression, rendering, and synchronization do not fit cleanly into traditional disciplinary boundaries.

The infrastructure gap further reinforces this fragmentation. Existing cloud, networking, and XR-related testbeds~\cite{fabric,han2023comic,illixr,cosmos,ara} provide important foundations, but significant gaps remain between general-purpose infrastructure and the requirements of immersive systems research. Participants emphasized the need for realistic XR workloads, heterogeneous devices, multi-user experimentation, real cloud infrastructure, programmable wireless networks, and human-in-the-loop evaluation capabilities.

This fragmentation slows progress because advances in one layer are often constrained by tradeoffs in another. A more compact neural representation may increase decoding latency, stronger privacy protections may reduce usability, and cloud offloading may save device power while increasing latency or reliability risks. As a result, immersive computing cannot be treated as a collection of independent problems in networking, systems, AI, graphics, or applications. Instead, the workshop framed immersive computing as a fundamentally cross-layer challenge requiring coordinated advances across representation, sensing, networking, computation, rendering, privacy, security, and human experience.
The central motivation for this report is therefore to articulate that shared challenge and identify a coordinated research agenda. 
The goal is not only to improve performance or realism, but also to make immersive computing scalable, trustworthy, accessible, reproducible, and deployable in real-world settings.

\section{A Multi-Layered Vision for Immersive Systems 
}
\begin{figure}
    \centering
    \includegraphics[width=0.9\linewidth]{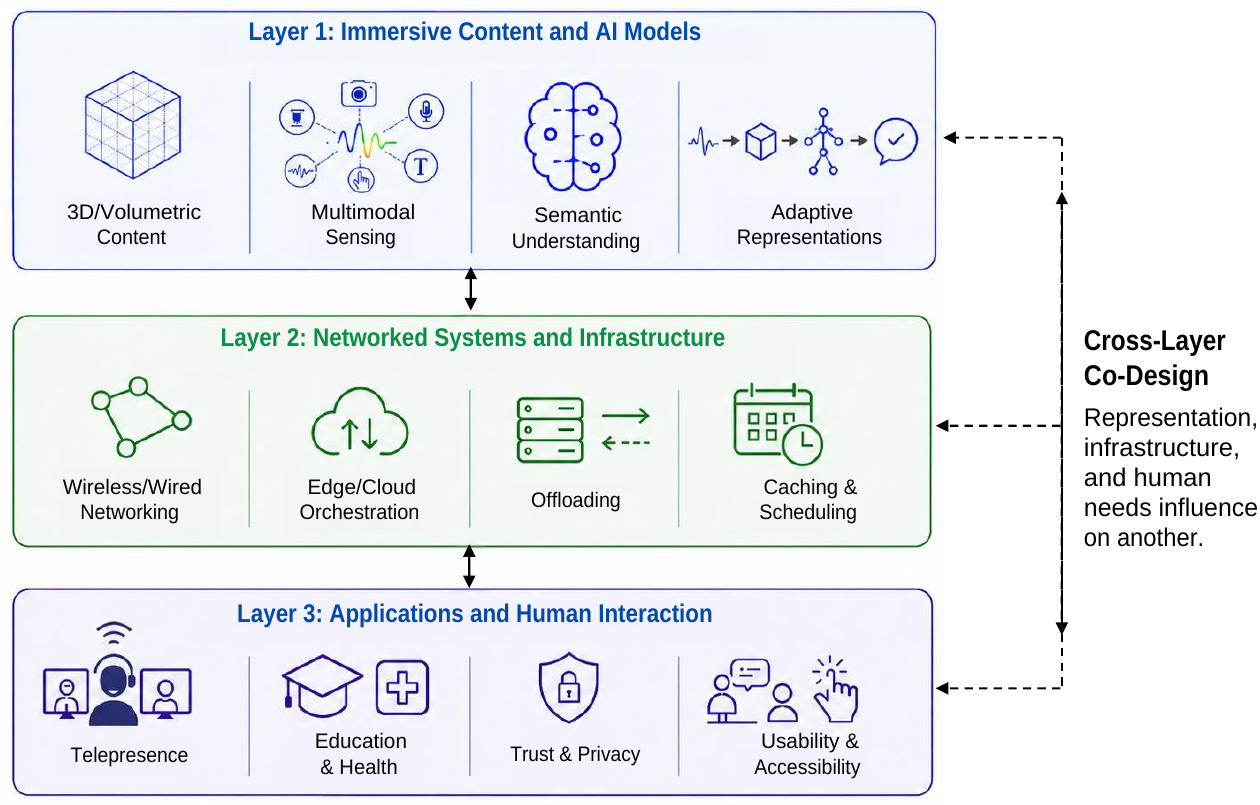}
    \caption{Multi-layer vision for immersive computing.
    }
    \label{fig:layered-vision}
\end{figure}

Immersive computing should be understood not as a single technology, but as a vertically integrated computing ecosystem in which content, intelligence, infrastructure, and human interaction are tightly coupled. A compelling immersive experience depends on the coordinated operation of content representations, AI models, distributed systems, networking, and human-centered interaction mechanisms.
The workshop therefore proposed a three-layer vision for immersive computing, organized around \textit{Immersive Content and AI Models}, \textit{Networked Systems and Infrastructure}, and \textit{Applications and Human Interaction}. As illustrated in Figure~\ref{fig:layered-vision}, these layers capture complementary concerns: Layer~1 focuses on representation and semantic understanding, Layer~2 focuses on runtime coordination and distributed infrastructure, and Layer~3 focuses on human experience, trust, and real-world deployment. Together, these layers provide a foundation for identifying shared research challenges and guiding future systems design.

\subsection{Layer 1: Immersive Content and AI Models}

The first layer concerns how immersive experiences are represented, interpreted, and adapted. Unlike traditional media, immersive systems must continuously integrate spatial geometry, multimodal sensing, contextual information, and semantic understanding into interactive environments. The \textit{central question} for this layer is: \textit{what representations best support scalable, intelligent, adaptive, and human-centered immersive interaction?}

Workshop discussions emphasized that current XR systems frequently capture and process substantially more information than applications ultimately require. This motivates adaptive representations that selectively encode geometry, semantics, intent, and contextual relevance rather than continuously transmitting raw sensory streams. Participants also identified the lack of standardized, hardware-friendly representations for emerging immersive media formats such as neural radiance field (NeRF)~\cite{mildenhall-cacm21} and 3D Gaussian splatting (3DGS)~\cite{kerbl20233d} as both a systems challenge and an opportunity to redesign immersive content pipelines around streamability, efficiency, and real-time interaction~\cite{wu2025nevo,wu2025advancing}.

AI and machine learning are increasingly central to this representation layer because they enable scene understanding, personalization, multimodal fusion, and context-aware interaction~\cite{afzal-cga25}. At the same time, the workshop emphasized that AI cannot be treated as a universal replacement for geometric or systems-driven methods. Future immersive systems will require hybrid approaches that balance semantic understanding, responsiveness, energy efficiency, privacy, and interpretability under real-world constraints.

The long-term vision for this layer is a family of adaptive immersive representations that can fluidly transition across levels of abstraction, from raw sensory signals to semantic objects and intent-level communication. Rather than asking only how to transmit more pixels, future systems must determine what information is truly necessary for a given interaction or task. These representation challenges naturally motivate the need for new runtime architectures and distributed infrastructure, which we discuss next.

\subsection{Layer 2: Networked Systems and Infrastructure}

The second layer concerns the runtime infrastructure that computes, coordinates, and delivers immersive experiences. This layer includes networking, distributed runtime systems, edge/cloud orchestration, synchronization, caching, and resource management. The \textit{central question} for this layer is: \textit{how should communication, computation, and storage be coordinated across devices, networks, edge systems, and cloud infrastructure to support real-time immersive interaction?}

Unlike traditional video delivery, immersive systems cannot rely heavily on buffering because responsiveness and synchronization directly shape user comfort and interaction quality~\cite{zhao2017estimating}. Workshop discussions emphasized that immersive infrastructure must dynamically adapt to changing network conditions, user behavior, device capabilities, and application requirements. As a result, the ``edge'' becomes a fluid systems abstraction rather than a fixed deployment location.

Participants also highlighted that immersive networking extends beyond packet transport. Future systems must continuously coordinate computation, communication, and resource management under dynamic constraints on the system and applications. These requirements point toward programmable, application-aware infrastructure capable of coordinating sensing, communication, rendering, and AI inference across heterogeneous environments.

The workshop further identified substantial infrastructure gaps in existing research platforms. While current networking and XR testbeds provide important foundations~\cite{fabric,han2023comic}, immersive computing research still lacks realistic multi-user workloads, heterogeneous device support, large-scale runtime experimentation, and reproducible human-in-the-loop evaluation.

The long-term vision for this layer is an immersive infrastructure fabric that exposes programmable interfaces for latency, reliability, energy efficiency, privacy, and QoE while supporting real-time, multi-user experiences at scale. Yet high-performance infrastructure alone is insufficient. Immersive systems must also satisfy the human, social, and application-level constraints that ultimately determine real-world adoption.

\subsection{Layer 3: Applications and Human Interaction}

The third layer concerns the human-facing goals of immersive systems: what users experience, what tasks are supported, and what forms of interaction become socially acceptable, trustworthy, and useful in practice. This layer includes applications such as telepresence, medical training, collaborative work, and human-AI interaction. The \textit{central question} for this layer is: \textit{what makes immersive computing genuinely valuable and trustworthy in real-world settings?}

Workshop discussions emphasized that immersive systems cannot be evaluated solely through technical metrics such as throughput or rendering quality. Long-term adoption depends equally on comfort, trust, accessibility, cognitive load, privacy, and social acceptance~\cite{Creed2024Inclusive}. Different application domains also impose fundamentally different requirements on latency, synchronization, spatial accuracy, reliability, and privacy, making one-size-fits-all immersive architectures unlikely to succeed.

Participants further stressed that immersive systems should be designed with users and domain experts rather than merely evaluated by them~\cite{jin2022will,Nafis2026Exploring}. Real-world deployment will require longitudinal studies, application-grounded evaluation, and interaction models that adapt to diverse users, environments, and accessibility needs. Security, privacy, and trust, therefore, become core challenges in interaction because immersive systems directly shape what users see, hear, and experience~\cite{lebeck2017securing,lebeck2018multiuser}.

The long-term vision for this layer is a human-centered immersive ecosystem that supports socially aware, inclusive, trustworthy, and adaptive interaction across diverse users and environments.
More broadly, immersive computing challenges the community to rethink networking, systems, and AI not as isolated layers, but as a unified infrastructure for shared human experience and intelligent digital interaction.

\begin{figure}
    \centering
    \includegraphics[width=1.0\linewidth]{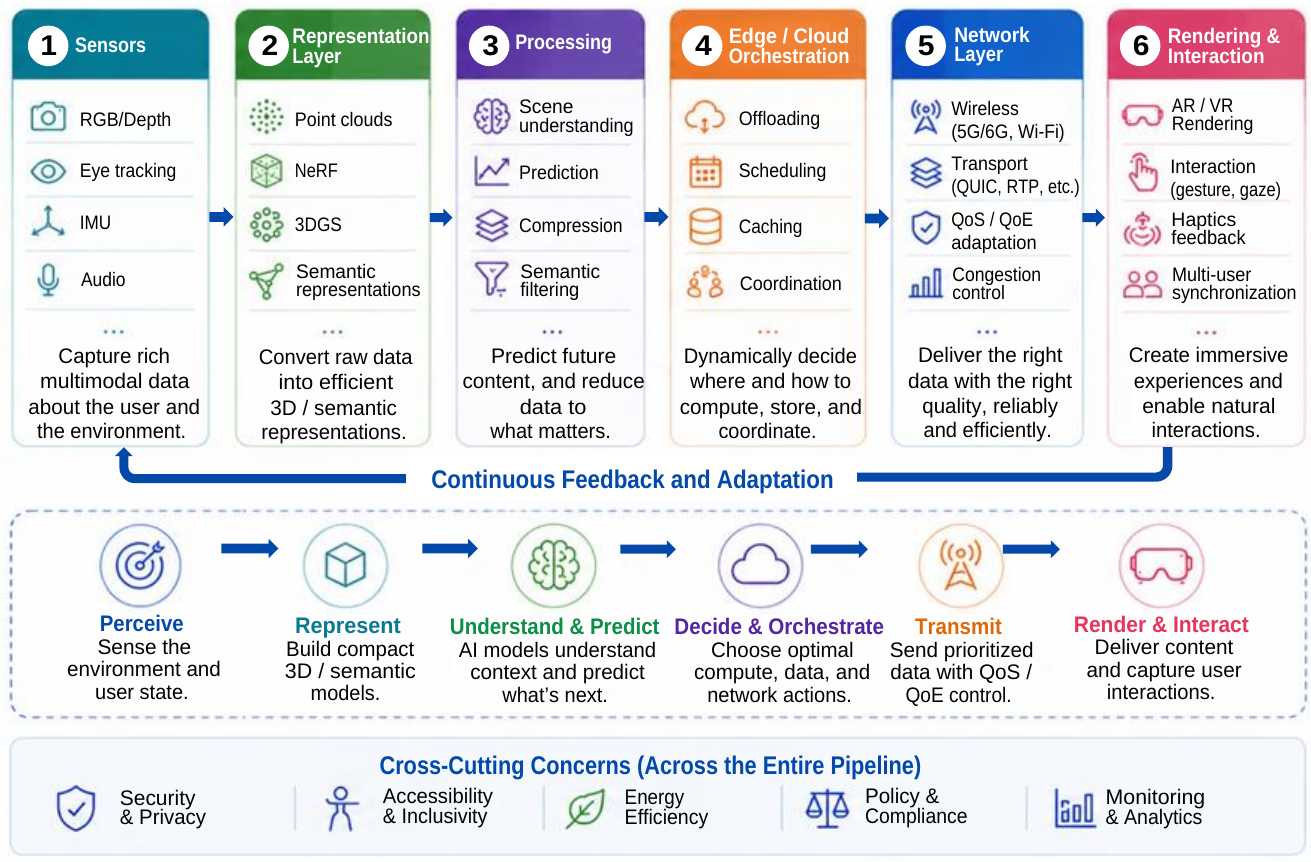}
    \caption{End-to-end immersive computing pipeline.
    }
    \label{fig:pipeline}
\end{figure}

\subsection{Cross-Layer Co-Design}

Immersive environments create continuous feedback loops across representation, communication, computation, rendering, privacy, and human interaction, making cross-layer coordination essential.
The workshop discussions highlighted several core design principles. First, immersive systems should prioritize information based on semantic importance, user attention, task relevance, and safety constraints rather than treating all data equally. Second, privacy, security, and trust must be integrated throughout the entire pipeline, from sensing and representation to networking, AI inference, and user interaction. Third, AI models must remain system-aware and human-aware, adapting to constraints on latency, energy, privacy, and interpretability rather than optimizing accuracy alone. Fourth, edge/cloud orchestration should be application-aware, since different immersive applications impose fundamentally different requirements on latency, reliability, synchronization, privacy, and spatial accuracy. Finally, evaluation must become both end-to-end and human-centered, incorporating not only system metrics such as latency and throughput, but also comfort, accessibility, trust, task success, and multi-user consistency.

Figure~\ref{fig:pipeline} illustrates this broader vision as a continuous and adaptive immersive computing loop. Human behavior, network conditions, privacy policies, application objectives, and system constraints continuously influence decisions about sensing, representation, communication, rendering, and interaction. The following section organizes the resulting technical challenges into six interconnected research domains spanning networking, systems, AI, security, human experience, and infrastructure. These challenge areas span all three layers of the architecture, with security and privacy extending across the full stack, and shared testbeds and infrastructure providing the foundation for reproducible, cross-layer immersive computing research.

\section{Key Challenges and Opportunities 
}

\begin{figure}
    \centering
    \includegraphics[width=0.9\linewidth]{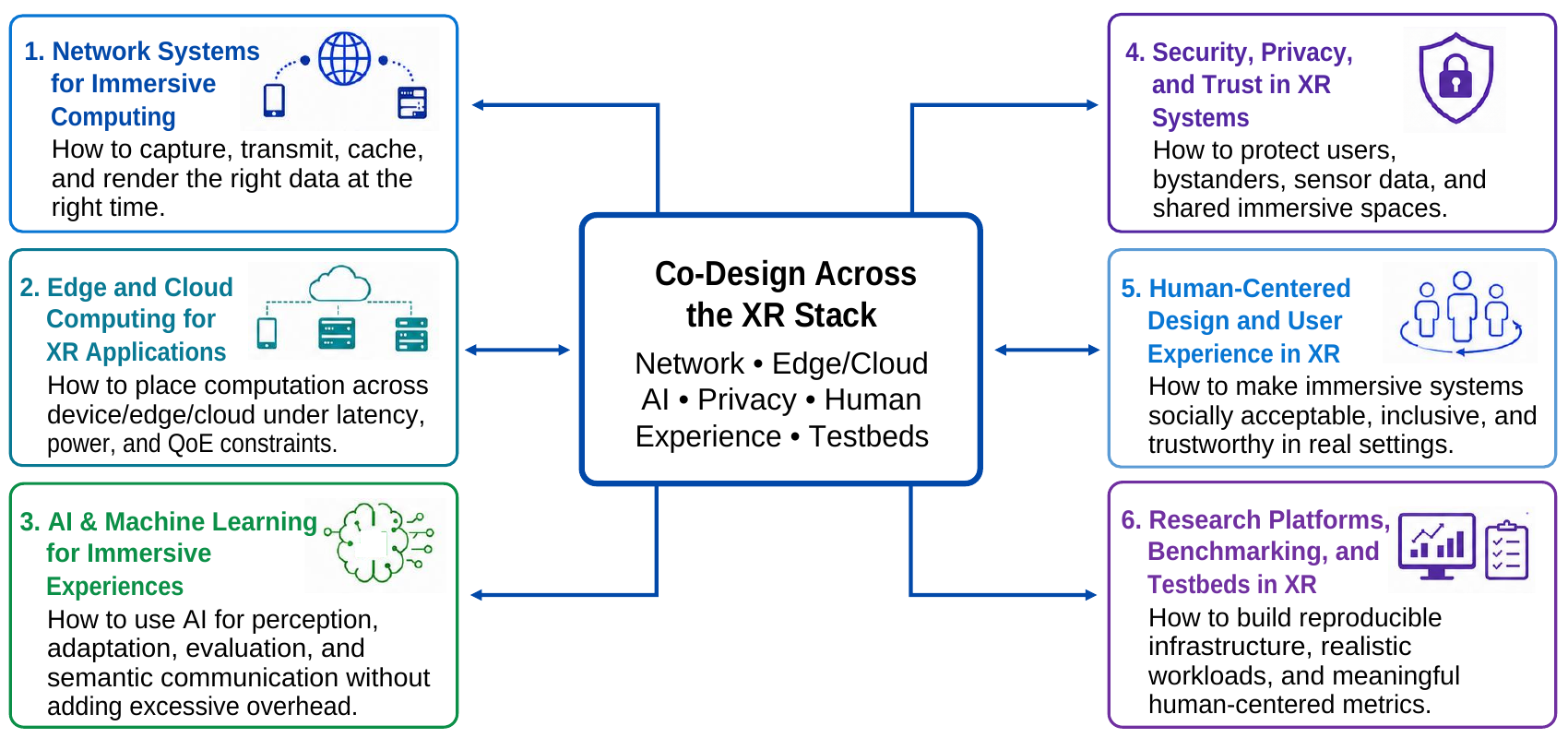}
    \caption{Challenge areas for end-to-end immersive computing, organized around six themes requiring cross-layer co-design.
    }
    \label{fig:challenges}
\end{figure}

Future XR systems require end-to-end co-design across networking, systems, sensing, AI, computing, security, privacy, human experience, and testbed infrastructure. As shown in Figure~\ref{fig:challenges}, the key challenge areas span networked systems, edge/cloud orchestration, AI-native immersive systems, security and privacy, human-centered XR design, and scalable platforms, benchmarks, and testbeds for reproducible immersive computing research. The following subsections summarize the key challenges and opportunities in each area.

\subsection{Networked Systems for Immersive Computing
}

Networked immersive systems must deliver interactive experiences on resource-constrained XR devices while meeting strict constraints on latency, energy, reliability, and privacy. Existing Internet architectures and media delivery pipelines were largely designed for passive content consumption and therefore struggle to support the real-time, adaptive, and multi-user requirements of immersive computing. Future XR systems require a broader shift from bulk data delivery toward context- and application-aware communication that adapts to latency, battery life, and user attention.

\paragraph{Challenges.}
A central challenge is the continuous generation and consumption of high-volume immersive data streams. XR systems must coordinate sensing, rendering, tracking, semantic understanding, and interaction under highly dynamic network and device conditions. Current architectures often rely on capture-first pipelines that acquire and process substantially more visual information than applications ultimately require, creating unnecessary computation, communication overhead, and privacy exposure.

A second challenge concerns the computation--communication tradeoff. Immersive systems must continuously decide where computation should occur, on-device, at the edge, or in the cloud, and what information should be transmitted, including raw sensory data, compressed representations, semantic features, or inference outputs~\cite{cheng2024magicstream,meng2022we}. On-device processing increases energy consumption, thermal load, and hardware complexity, while remote offloading introduces latency, reliability, and privacy risks. These tradeoffs vary significantly across immersive applications and operating environments.

Another challenge is the lack of standardized, hardware-efficient representations and delivery mechanisms for immersive media. Emerging representations such as NeRF~\cite{mildenhall-cacm21} and 3DGS~\cite{kerbl20233d} introduce new demands on streaming, caching, rendering, and synchronization that are poorly matched to existing video-centric networking and compression pipelines~\cite{han2020vivo,lee2020groot}. Future systems must support scalable and adaptive immersive representations under tight resource constraints.

\paragraph{Opportunities.}
A major opportunity is to move beyond raw data transmission toward selective and semantic communication~\cite{cheng2024magicstream}. Rather than treating all pixels or sensory streams equally, immersive systems can prioritize information based on user attention, object importance, task relevance, safety constraints, and application context. Selective delivery, adaptive fidelity, and intelligent caching of structured 3D representations can significantly reduce bandwidth demand while improving responsiveness and scalability.

Another opportunity is the development of hybrid sensing pipelines that combine visual information with lower-overhead contextual signals. Instead of relying exclusively on camera-centric sensing, future systems can selectively fuse complementary localization and wireless sensing modalities with visual data to reduce sensing overhead, improve privacy, and maintain sufficient spatial awareness for immersive interaction~\cite{Kamari2026View}. These approaches point toward more adaptive and resource-aware sensing architectures.

Future immersive networking systems should also become application-aware and programmable. Different XR applications impose fundamentally different requirements on latency, synchronization, privacy, fidelity, and scalability. Supporting these diverse requirements will require new abstractions for naming, discovery, synchronization, and distributed resource coordination across immersive environments.

\paragraph{Vision.}
The long-term vision is an immersive networking architecture that treats communication, computation, sensing, and semantic understanding as a unified adaptive system. Rather than optimizing raw throughput alone, future immersive networks should dynamically determine what information matters, where computation belongs, and how resources should be coordinated to support scalable, trustworthy, and human-centered immersive experiences.

\subsection{Real-Time Edge–Cloud Orchestration
}

XR orchestration is fundamentally a real-time systems problem. Immersive applications must continuously coordinate computation, communication, storage, and rendering across devices, edge systems, and cloud infrastructure under strict constraints involving latency, energy, reliability, privacy, and user experience. Existing cloud and mobile computing architectures were not designed for persistent, interactive, and spatially grounded workloads, motivating a shift toward adaptive and application-aware runtime systems.

\paragraph{Challenges.}
A central challenge is determining where immersive computation should occur. XR systems must dynamically partition workloads across resource-constrained devices, nearby edge resources, and remote cloud infrastructure while balancing latency, energy consumption, reliability, and privacy~\cite{satyanarayanan2009case,shi2016edge}. These decisions become especially difficult because immersive applications continuously generate time-sensitive state, including scene representations, user motion, interaction context, and AI inference outputs that cannot tolerate stale synchronization or excessive buffering.

Another challenge is the lack of principled orchestration frameworks and evaluation models for immersive systems. Current XR runtimes often rely on heuristic offloading and adaptation strategies, while existing QoE metrics remain insufficient for reasoning about the complex tradeoffs among latency, interaction stability, visual quality, comfort, and energy efficiency. As immersive applications diversify, static deployment assumptions and fixed orchestration policies become increasingly ineffective.

A further challenge is the heterogeneity of immersive applications and deployment environments. Different XR workloads impose fundamentally different requirements on privacy, synchronization, reliability, and infrastructure locality. Systems designed for telepresence/teleoperation, medical assistance, or collaborative training may therefore require substantially different runtime behaviors and orchestration strategies.

\paragraph{Opportunities.}
A major opportunity is the development of dynamic and partitionable XR runtimes that continuously adapt computation placement and resource coordination in response to changing system conditions. Rather than optimizing throughput alone, future systems should jointly optimize latency, responsiveness, reliability, energy efficiency, and human experience under application-specific constraints.

Another opportunity is the emergence of programmable edge infrastructure for immersive computing. Future runtime systems may integrate distributed AI inference, rendering, sensing, and communication into coordinated service pipelines spanning devices, networks, edge systems, and cloud platforms. This direction connects immersive computing with broader advances in programmable networking, distributed AI systems, and real-time infrastructure orchestration~\cite{fabric}.

A third opportunity is predictive and proactive orchestration for immersive environments. Future XR systems may increasingly anticipate user motion, interaction patterns, network variability, and workload dynamics before performance degradation occurs. By combining predictive scheduling, proactive caching, workload migration, and adaptive prefetching, immersive platforms can reduce latency spikes, minimize service interruption, and improve interaction stability under highly dynamic conditions. Such approaches may become particularly important for mobile multi-user XR, large-scale collaborative environments, and geographically distributed immersive applications.

\paragraph{Vision.}
The long-term vision is an immersive runtime fabric that continuously determines what to compute, where to compute it, and how to coordinate resources across devices, networks, edge systems, and cloud infrastructure in real time. Achieving this vision requires moving beyond static offloading heuristics toward adaptive, programmable, and application-aware orchestration for immersive computing.

\subsection{AI-Native Immersive Experiences
}

AI is becoming a foundational component of immersive systems because it enables perception, adaptation, semantic understanding, and interactive behavior at scales that traditional pipelines cannot easily support. At the same time, immersive environments impose stringent constraints involving latency, energy, privacy, reliability, and human experience, requiring intelligent adaptation to be designed as part of the broader systems stack rather than as an isolated inference layer.

\paragraph{Challenges.}
A central challenge is integrating learned models into latency-sensitive immersive pipelines. Many models prioritize output quality over predictable execution time, making them difficult to deploy in interactive XR environments where delays directly affect responsiveness, comfort, and usability~\cite{guan2022deepmix}. Future systems must therefore determine when learned approaches provide meaningful benefits, when lightweight geometric methods remain preferable, and how hybrid pipelines can balance semantic understanding with real-time performance.

Another challenge concerns combining semantic reasoning with physically grounded immersive systems. Traditional XR pipelines rely heavily on geometry, tracking, calibration, and explicit scene reconstruction, while learned approaches offer adaptability and contextual understanding~\cite{hou2025falcon}. Integrating these techniques remains difficult because immersive systems must operate reliably across diverse users, environments, and interaction patterns while preserving spatial consistency and predictable behavior.

A further challenge is evaluation and trustworthiness. Intelligent immersive systems may help interpret content and automate adaptation, but they must also account for human-centered properties such as comfort, timing, spatial coherence, task success, and the quality of social interaction. At the same time, these systems introduce new risks involving privacy exposure, manipulated environments, adversarial inputs, and behavioral inference from motion, gaze, and physiological signals~\cite{shi2025your}.

\paragraph{Opportunities.}
One major opportunity is adaptive perception and interaction for immersive environments. Future systems can use contextual understanding of user behavior, spatial interaction, and application intent to optimize rendering, sensing, communication, and resource allocation in real time. Rather than statically configuring immersive pipelines, immersive platforms can dynamically adapt fidelity, interaction behavior, and computation placement to changing system and user conditions.

A second opportunity is the development of context-aware AI agents for immersive environments~\cite{chen2024immersive}. Such agents could understand user goals, physical context, interaction history, and application intent to assist with navigation, education, collaboration, and embodied interaction. Future immersive agents may also support persistent spatial memory, multimodal reasoning, and collaborative human--AI interaction within shared immersive environments.

A third opportunity is the development of explainable and physically grounded AI for XR systems. Many current AI models operate as black boxes, making it difficult to understand or validate their decisions in interactive and safety-critical immersive environments. Future XR systems may benefit from interpretable AI models that incorporate geometric constraints, physical knowledge, and human-understandable reasoning into decision-making. Such approaches can improve transparency, debugging, adaptive control, and user trust while enabling immersive systems that remain responsive to real-world physical and spatial constraints.

\paragraph{Vision.}
The long-term vision is not simply to insert larger models into XR pipelines, but to co-design intelligence, systems, networking, and human interaction as a unified immersive computing stack. Future immersive systems should selectively and efficiently use learned adaptation, enabling richer interaction and semantic understanding while preserving responsiveness, trustworthiness, and real-time human experience.

\subsection{Privacy-Preserving Immersive Environments
}

Security and privacy are foundational challenges for immersive computing because XR systems directly shape what users see, hear, feel, and believe. Unlike traditional computing platforms, immersive systems continuously sense users and surrounding environments while simultaneously influencing perception and behavior through interactive visual, audio, and haptic feedback~\cite{cheng2023exploring}. These capabilities significantly expand the attack surface of computing systems and require a broader view of trust that spans sensing, interaction, rendering, networking, and human experience.

\paragraph{Challenges.}
A central challenge is the unprecedented volume and sensitivity of immersive telemetry. XR devices continuously collect spatial, behavioral, and physiological information through tracking, environmental sensing, and multimodal interaction~\cite{wu2023privacy,zhang2023facereader}. These signals can reveal identity, habits, emotional state, private spaces, social relationships, and other sensitive contextual information, often beyond what users or bystanders realize is being inferred.

Another challenge is that immersive systems can directly manipulate perception. Malicious or compromised XR applications may alter virtual content, distort spatial understanding, inject misleading information, or influence user behavior in ways that extend beyond conventional data security concerns~\cite{cheng2023exploring}. In shared immersive environments, these risks become even more complex because avatars, virtual objects, and collaborative interactions shape how users interpret both digital and physical reality.

A further challenge is the lack of an open and reproducible security infrastructure for immersive systems. Many XR platforms remain proprietary and difficult to inspect, limiting independent evaluation of defenses, privacy mechanisms, and safety guarantees~\cite{shoaib2025realitycheck}. At the same time, immersive computing introduces difficult tradeoffs among privacy, utility, latency, usability, and human experience that existing security models do not adequately capture.

\paragraph{Opportunities.}
A major opportunity is to redesign XR security and privacy as end-to-end system properties rather than application-level add-ons. Future immersive platforms should integrate privacy-aware sensing, fine-grained access control, secure rendering, trusted interaction mechanisms, and transparent telemetry management directly into runtime systems and infrastructure~\cite{kim2023erebus}. These protections must remain compatible with the responsiveness and usability requirements of immersive interaction.

Another opportunity is to develop trust-aware immersive systems that model trust relationships among users, platforms, applications, collaborators, and physical environments. Rather than treating trust as a binary property, future systems should support adaptive policies governing who can access data, modify shared environments, or influence user perception in collaborative immersive spaces~\cite{lebeck2018multiuser}.

Immersive environments also create opportunities for security research itself. XR platforms can support realistic simulations of cyber-physical threats, security training, social engineering studies, and human-centered security evaluation in controlled and reproducible settings. These capabilities may ultimately make security research more interactive, experiential, and accessible.

\paragraph{Vision.}
The long-term vision is trustworthy immersive computing in which sensing, interaction, rendering, and communication are designed around human safety, transparency, and control from the outset. Future XR systems must protect not only data confidentiality, but also perception, cognition, physical safety, and social trust within shared immersive environments.

\subsection{Human-Centered Design and User Experience 
}

Immersive computing ultimately succeeds or fails based on how people experience it in real-world settings. Beyond technical performance, XR systems must be socially acceptable, trustworthy, inclusive, and genuinely useful for diverse users and across diverse environments. Existing XR research still relies heavily on short-term laboratory studies, motivating a broader shift toward longitudinal, application-grounded, and human-centered evaluation.

\paragraph{Challenges.}
A central challenge is social legitimacy. Lightweight and wearable XR devices may reduce barriers to adoption, but immersive systems also raise concerns about visibility, surveillance, distraction, and social discomfort in shared environments. These concerns affect not only users but also bystanders who may be recorded or influenced without clear awareness or consent.

Another challenge is determining when immersive interaction provides meaningful task-level value. While XR can support spatial understanding, embodied learning, and remote collaboration, not every task benefits from immersive presentation. In certain cases, immersive environments may introduce additional cognitive load, physical discomfort, setup complexity, or distraction without improving user outcomes~\cite{jin2022will,jin2024virtual}. These tradeoffs are particularly important in education and workplace settings, where usability and attention directly affect effectiveness.

A further challenge is the design of inclusive and ethical systems. Immersive systems must accommodate users with different cognitive, developmental, and social needs while avoiding unintended harms such as sensory overload, exclusion, dependency, or unequal access. As XR becomes more integrated into learning, communication, and care environments, designers must carefully balance personalization, engagement, privacy, accessibility, and long-term well-being.

\paragraph{Opportunities.}
A major opportunity is to embed stakeholder involvement directly into the design of immersive systems. Teachers, caregivers, clinicians, accessibility communities, and other domain experts can help ensure that XR systems address real needs and deployment constraints rather than technology-driven use cases~\cite{jin2022will,Nafis2026Exploring}. This direction encourages application-grounded design practices that connect technical innovation with meaningful human benefit.

Another opportunity is the development of modular and adaptable immersive platforms. Flexible interaction models, reusable infrastructure, and multimodal interfaces can allow the same XR platform to support diverse use cases such as remote assistance, collaborative learning, and human-AI interaction~\cite{cheng2025hello}. These systems may also enable AI-assisted interaction mechanisms that reduce user burden through contextual guidance, translation, summarization, and adaptive support~\cite{cheng2025hello}.

Longitudinal and real-world evaluation also represents a major opportunity for the field. Long-term deployments can reveal how privacy expectations, social norms, accessibility needs, and interaction behavior evolve over time in shared immersive environments. Such studies are essential for understanding how immersive systems fit into everyday life beyond controlled laboratory settings.

\paragraph{Vision.}
The long-term vision is a human-centered immersive ecosystem that is not only technically capable but also socially legitimate, inclusive, and human-centered. Future XR systems should adapt to diverse users, environments, and social contexts while supporting interaction models that enhance human capability without increasing cognitive burden or reducing trust.

\subsection{Scalable Infrastructure, Testbeds, and Open Ecosystems 
}
Progress in immersive computing depends on shared infrastructure capable of supporting realistic experiments involving devices, networks, edge/cloud systems, applications, and human users. While existing networking and XR platforms provide valuable foundations, current infrastructure remains insufficient for large-scale, end-to-end research on immersive systems. Future XR ecosystems require a broader shift from isolated prototypes toward reproducible, application-grounded, and cross-layer experimentation.

\paragraph{Challenges.}
A central challenge is the gap between general-purpose infrastructure and XR-specific environments. Existing networking testbeds provide strong support for wireless, cloud, and distributed systems experimentation~\cite{fabric,cosmos}, while XR-focused platforms support immersive runtimes, rendering, and interaction~\cite{han2023comic,illixr}. However, these systems remain only loosely integrated, limiting the ability to study immersive applications across sensing, networking, edge/cloud orchestration, rendering, and human interaction within a unified experimental framework.

Another challenge is the lack of realistic workloads, user traces, and evaluation methodologies for immersive systems. XR applications involve highly dynamic user behavior, spatial interaction, multimodal sensing, and strict real-time requirements that are difficult to reproduce through conventional simulations or static benchmarks. Existing evaluation practices also struggle to connect low-level system metrics with higher-level measures of immersion, comfort, accessibility, task performance, and multi-user interaction quality.

A further challenge is anticipating future immersive systems using today's infrastructure. Many current XR platforms remain limited in scale, realism, and reliability, making it difficult to evaluate next-generation immersive applications involving digital twins, persistent virtual environments, large-scale collaboration, or AI-native interaction.

\paragraph{Opportunities.}
A major opportunity is to build integrated XR research ecosystems that combine networking infrastructure, edge/cloud orchestration, immersive runtimes, programmable wireless systems, and human-centered evaluation into shared experimental platforms. Existing infrastructures provide a strong starting point for creating end-to-end immersive computing environments that support reproducible, scalable cross-layer experimentation~\cite{han2023comic}.

Another opportunity is the use of digital twins and model-based experimentation for immersive systems research. Virtual cities, industrial environments, and collaborative digital spaces can enable scalable evaluation of networking, AI, interaction, and orchestration strategies under realistic but controllable conditions. These approaches may also reduce deployment cost and improve reproducibility for complex immersive workloads.

The community also has an opportunity to establish shared repositories for immersive workloads, datasets, user traces, and evaluation artifacts. Open and reusable resources can lower barriers to entry, improve comparability across systems, and accelerate cumulative progress in immersive computing research.

\paragraph{Vision.}
The long-term vision is an open and interoperable immersive computing research ecosystem that integrates infrastructure, workloads, datasets, benchmarks, digital twins, and human-centered evaluation into a shared foundation for the community. Such infrastructure would enable immersive computing research to move beyond isolated demonstrations toward reproducible, scalable, and application-grounded systems science.

\paragraph{Key Takeaways.}
\begin{itemize}
    \item Immersive computing is fundamentally a cross-layer systems problem spanning sensing, representation, networking, computation, AI, privacy, and human interaction.
    
    \item Future XR systems must move beyond bandwidth-centric design toward adaptive, semantic, and context-aware architectures that optimize what information is sensed, transmitted, computed, and rendered.
    
    \item Privacy, trustworthiness, accessibility, and social acceptability are central design requirements rather than secondary concerns for immersive systems.
    
    \item Scalable progress requires shared infrastructure, reproducible evaluation methodologies, interoperable platforms, and closer collaboration across academia, industry, government, and application domains.
\end{itemize}

\section{Key Research Questions 
}
This section synthesizes core research questions and answers to clarify both the research challenges and the community's emerging direction.

\paragraph{Q1: What is the fundamental bottleneck: networking, compute, or representation?}

\textbf{A1:}
There is no single bottleneck. Immersive systems are constrained by the interaction among network bandwidth, latency, device power, compute placement, sensing, compression, and content representation. A better network alone will not solve the problem if the representation is inefficient; a better AI model may not help if it adds latency; and more on-device compute may reduce offloading but increase heat, weight, and battery drain. The key research direction is therefore cross-layer co-design across representation, computation, communication, and human experience.

\paragraph{Q2: Can semantic communication replace raw data streaming?}

\textbf{A2:}
Semantic communication can reduce the need to transmit raw pixels, point clouds, or sensor streams, but it is unlikely to fully replace raw data in all scenarios. Instead, future systems will likely use hybrid approaches: raw or high-fidelity data for safety-critical or visually complex content, and semantic representations for intent, object state, scene structure, motion, and task-relevant context. This direction aligns with the broader observation that not all pixels are equally important, and that immersive systems should prioritize information based on user attention, application context, and task relevance. Over time, semantic communication may significantly reduce bandwidth demand while improving scalability and responsiveness.

\paragraph{Q3: How can immersive systems scale to 100+ concurrent users?}

\textbf{A3: }
Scaling to large multi-user environments requires more than increasing server capacity. Systems must manage shared state, synchronize users, reuse common views or representations, cache and prefetch 3D content, and decide which updates are essential for each user. Efficient mechanisms for shared-state synchronization, view sharing, adaptive caching, and interest management will become increasingly important as user populations grow. Large-scale XR also requires application-aware networking, multi-user QoE metrics, and edge/cloud orchestration that can adapt to group behavior, user mobility, and geographic distribution.

\paragraph{Q4: What is the ``ideal'' immersive system architecture?}

\textbf{A4: }
The ideal architecture is not a fixed device-edge-cloud pipeline. It should be dynamic, partitionable, and application-aware. Some workloads may run on glasses, some on a phone or compute puck, some at a nearby edge server or base station, and some in the cloud. The architecture should continuously decide what to sense, compute, transmit, cache, and render based on latency, power, privacy, safety, and application needs.

\paragraph{Q5: How should QoE be measured beyond video metrics?}

\textbf{A5: }
XR QoE must go beyond frame rate, bitrate, and visual fidelity. It should include responsiveness, interaction stability, cybersickness, cognitive load, comfort, task success, accessibility, privacy expectations, social acceptability, and multi-user consistency. Unlike traditional video streaming, immersive systems involve active user motion, spatial perception, real-time feedback, and social interaction. New metrics and models are needed to capture these human-centered dimensions and connect system behavior to real user experience.

\paragraph{Q6: How can XR systems protect bystanders as well as users?}

\textbf{A6: }
Bystander privacy is a defining challenge for AR and smartglasses. Devices may capture people who did not consent, and visual data can reveal their identities, emotions, behaviors, or other sensitive attributes. Protection mechanisms may include visible recording signals, no-recording zones, wireless beacons, on-device filtering, privacy-aware APIs, and policies that let public or private spaces communicate privacy expectations to devices. Prior research has also demonstrated gaze- and interaction-aware filtering techniques that can blur or suppress bystander information in real time when users are not actively interacting with it.

\paragraph{Q7: What shared infrastructure does the research community need?}

\textbf{A7: }
The community needs XR testbeds, benchmarks, datasets, and open-source platforms. Existing networking and cloud testbeds provide a foundation, but XR research also requires real users, heterogeneous devices, real-time workloads, edge/cloud support, human-experience metrics, and reproducible multi-user scenarios. Shared infrastructure is essential for moving from isolated prototypes to comparable and reusable research results. Standardized datasets, traces, and evaluation methodologies can further accelerate reproducibility and community-wide collaboration.

\paragraph{Q8: How can immersive systems become socially acceptable and trustworthy?}

\textbf{A8: }
Technical performance alone will not guarantee adoption. Immersive systems must fit into everyday life, respect social norms, support vulnerable populations, and make privacy and recording behavior understandable. Human-centered design should be embedded from the beginning through stakeholder involvement, longitudinal studies, modular systems, and policies that align with real environments such as schools, hospitals, workplaces, and elder-care facilities. Long-term success will depend not only on technical capability but also on public trust, accessibility, and social legitimacy.

\section{Important Future Trends 
}

\begin{figure}
    \centering
    \includegraphics[width=1.0\linewidth]{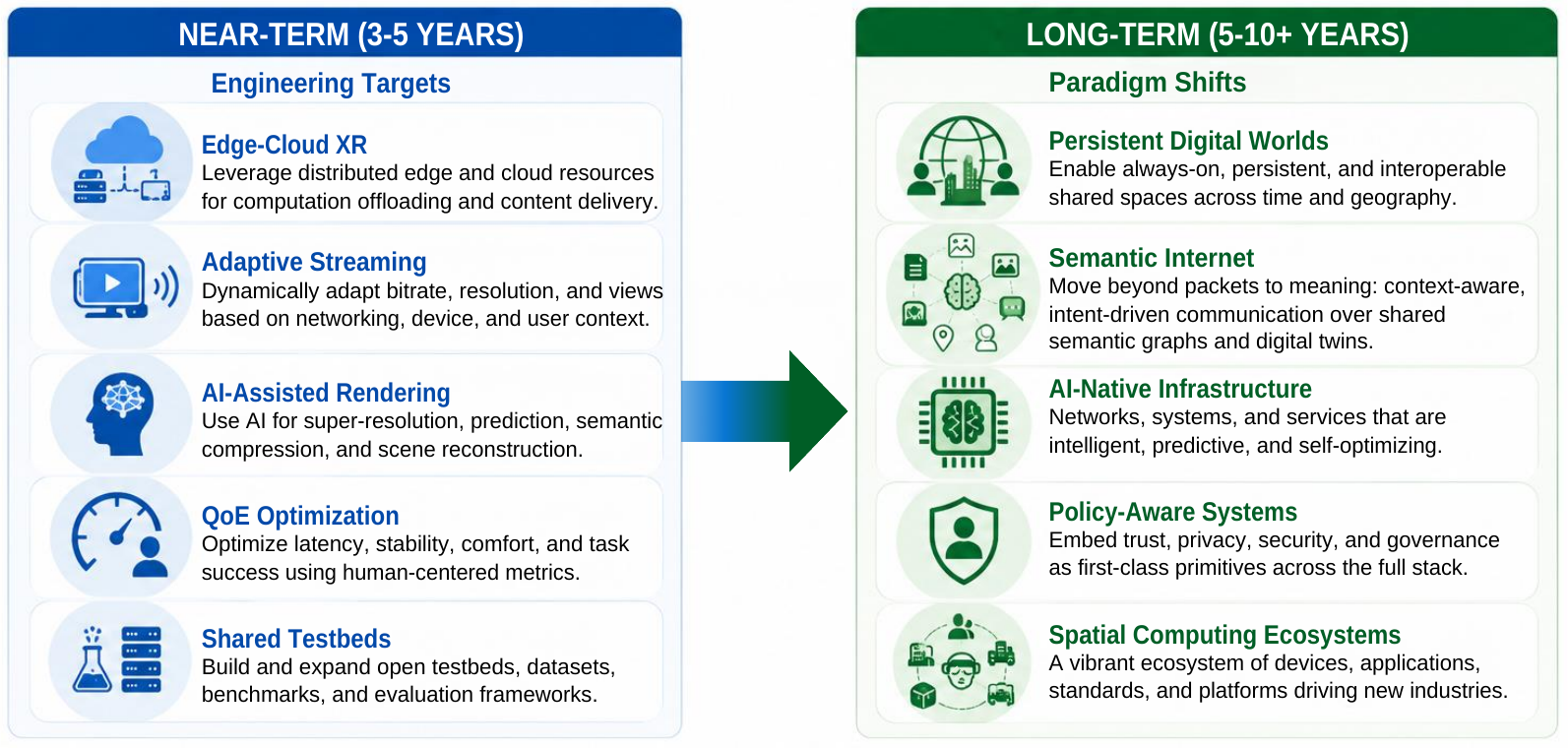}
    \caption{Evolution of Immersive Computing: From engineering targets to paradigm shifts.}
    \label{fig:trends}
\end{figure}

Synthesizing insights from the workshop, we anticipate that the trajectory of immersive computing will unfold through two complementary phases. 
As shown in Figure~\ref{fig:trends}, in the near term (3--5 years), progress will primarily be driven by concrete engineering targets, including the integration of existing XR hardware with edge-cloud infrastructure~\cite{han2023comic}, on-device and cloud-hosted AI models~\cite{Du2023Rapsai}, and next-generation networking systems into highly adaptive, cross-layer pipelines~\cite{zhang2022m5}. These efforts focus on improving scalability, responsiveness, efficiency, usability, and deployability under real-world constraints. Looking further ahead (5--10 years), immersive computing may evolve through broader paradigm shifts toward semantic, intent-driven, and AI-native infrastructures~\cite{han2026ai,Lee2025Sensible}, wherein human-centered interaction, spatial sensing, dynamic rendering, and network transport are co-designed as a unified, intelligent ecosystem. 
The longer-term directions are necessarily more exploratory, reflecting the interdisciplinary advances still required and the inherent uncertainty of a decade-long horizon.

\subsection{Near-Term Trajectories: 3--5 Years}

\paragraph{Edge-assisted XR.}
To accommodate the lightweight form-factor and thermal constraints of wearable displays, near-term XR systems will increasingly rely on opportunistic edge resources to support latency-sensitive tasks~\cite{Du2020DepthLab,zhang2018jaguar}. 
Crucially, the ``edge'' will evolve from a single mobile phone into a fluid computing continuum. Workloads will be dynamically distributed across smartglasses, mobile phones, Wi-Fi access points, cellular base stations, local edge devices such as cloudlets~\cite{satyanarayanan2009case}, and private and commercial cloud servers. The primary systems challenge lies in developing adaptive runtime schedulers that optimize task placement under fluctuating network latency, power budgets, user mobility, and strict privacy requirements~\cite{satya2017edge, sukhmani2019edge}.

\paragraph{Hybrid local--remote rendering.}
The dichotomy between local and remote rendering will blur into continuous, hybrid pipelines. Lightweight wearables will perform minimal local computation, such as late latching or asynchronous reprojection~\cite{lai2017furion}, while nearby devices or edge servers handle computationally intensive tasks, such as volumetric rendering, scene reconstruction, or generative AI inference~\cite{boos2016flashback,cuervo2015kahawai}. This split-rendering approach preserves local responsiveness while drastically elevating visual fidelity. Consequently, future mobile systems must dynamically partition perception and rendering workloads across the device-edge-cloud continuum while guaranteeing visual consistency and interaction stability under volatile network conditions~\cite{guan2022deepmix}.

\paragraph{AI-assisted compression and reconstruction.}
AI will fundamentally alter the cost structure of immersive media transport. Learned neural representations can help to compress, predict, and reconstruct high-fidelity volumetric content~\cite{wu2025nevo}, dynamic point clouds~\cite{kamari2023mmsv}, 3DGS~\cite{Hu2025Thing2Reality}, multiview video~\cite{du2018montage4d}, and complex light fields~\cite{meng20203d}. Furthermore, AI-driven pipelines will enable selective transmission strategies, such as streaming only salient keyframes~\cite{prada2016motion}, semantic feature maps, or coarse base layers that are hallucinated and refined entirely on the client side. However, deploying these models requires rigorous co-design; the latency and energy overhead of on-device neural inference must not negate the bandwidth savings~\cite{huynh2017deepmon}.

\paragraph{Application-aware sensing and delivery.}
The current paradigm of capturing and transmitting raw, unfiltered sensor data is unsustainable. Instead, sensing and delivery will become highly selective and deeply contextualized. For instance, the data requirements for surgical telementoring~\cite{gasques2021artemis, groves2025assessing}, remote scientific visualization~\cite{Nafis2026Exploring}, and casual social XR~\cite{cheng2024first, Du2019Geollery} vary widely in terms of visual acuity, spatial tracking, haptic feedback, and semantic metadata. Future architectures will proactively filter sensor streams, asking what minimum data is required for a specific task, what can be aggressively pruned, and what must remain localized to preserve user privacy~\cite{wu2022privacy}.

\paragraph{Human-centered QoE models.}
The community must transcend traditional video-streaming metrics (e.g., bitrate, resolution) to accurately evaluate immersive QoE. Near-term research will formulate holistic metrics that bridge objective system performance with subjective human factors. This includes quantifying motion-to-photon latency~\cite{zhao2017estimating}, visual stability, cybersickness~\cite{krokos2022quantifying}, cognitive load~\cite{Yoghourdjian2021Scalability}, task completion efficiency, and even physiological responses via electroencephalogram (EEG)~\cite{krokos2022quantifying}. Furthermore, socio-technical metrics such as privacy expectations~\cite{Liu2024Human}, accessibility~\cite{Jain2022ProtoSound}, and social acceptability~\cite{Creed2024Inclusive} will become foundational for benchmarking edge orchestration and adaptive streaming algorithms.

\paragraph{Privacy-by-design XR platforms.}
As XR headsets and smartglasses transition to ubiquitous daily use, privacy-aware platforms will become increasingly important. Near-term systems must advance bystander privacy protections, continuous motion-based authentication, real-time on-device filtering, verifiable recording indicators, and fine-grained, context-aware sensor APIs~\cite{david2024understanding,wu2025pipe}. Privacy can no longer be an afterthought; it must be treated as an end-to-end invariant across sensing, computation, transport, and rendering~\cite{Corbett2025ShouldAR,shi2025your}.

\subsection{Long-Term Trends: 5--10 Years}

\paragraph{Semantic and intent-driven networking.}
Over the next decade, immersive communication networks will undergo a paradigm shift: from transmitting raw pixels and explicit 3D geometries to communicating semantic meaning, human intent, and task-relevant state changes~\cite{cheng2024magicstream,xia2024sss}. By streaming object semantics, forecasted kinematics, and interaction intents, systems will drastically reduce payload sizes and mask network latency, enabling massively scalable immersive applications. The defining research challenge will be establishing the theoretical limits of semantic fidelity: knowing precisely when high-level semantic abstractions suffice and when raw, high-fidelity data is strictly necessary.

\paragraph{Multi-agent immersive systems.}
Future spatial environments will be populated not just by human avatars but also by proactive~\cite{Lee2025Sensible}, spatially aware~\cite {Liu2026AgentHands} AI agents that collaborate seamlessly across the physical-digital divide. Functioning as tutors, translators, collaborative partners, or safety monitors, these agents will require deep multimodal comprehension that fuses speech, micro-gestures, gaze, biometrics, spatial layouts, and historical context. This evolution will unlock profound new paradigms in human–AI interaction, while simultaneously surfacing complex challenges in algorithmic transparency, shared control, and verifiable trust~\cite{Hu2025DialogLab}.

\paragraph{Persistent shared digital worlds.}
XR systems will mature to support persistent, digital-physical spaces that evolve continuously across asynchronous user sessions and distributed geographic locations. These shared spatial computing environments will anchor next-generation education~\cite{cheng2024understanding}, therapeutic healthcare~\cite{white2026context}, surgical training~\cite{groves2025assessing}, and industrial digital twins~\cite{Du2019Geollery}. Realizing this vision requires massive-scale state synchronization, distributed access control, cryptographic provenance, and sophisticated conflict-resolution mechanisms for spatial data~\cite{zhao2022metaverse}. Open research challenges include maintaining low-latency consistency across geographically distributed users while preserving privacy, trust, and semantic coherence in continuously evolving virtual-physical environments.

\paragraph{AI-native immersive Internet.}
Ultimately, immersive computing will converge into an AI-native Internet~\cite{han2026ai}. Here, physical sensing, network transport, rendering, and interaction are continuously mediated and optimized by foundation models. These architectures will transcend traditional content streaming~\cite{guan2023metastream}; they will anticipate human context, autonomously personalize spatial representations, and orchestrate compute resources across the edge-cloud continuum. Achieving this requires a new class of AI models that are inherently latency-aware, privacy-preserving, interpretable, and resilient to the chaotic dynamics of the physical world~\cite{hou2024towards}.

\paragraph{Policy-aware and trustworthy infrastructure.}
As immersive technologies permeate public infrastructure and personal spaces, verifiable trust and policy enforcement must be integrated into the network and systems architecture. Future systems may support privacy zones, no-recording signals~\cite{bukhari2025rethinking}, decentralized trust models, zero-trust security~\cite{cheng2023toward}, fine-grained data governance, and transparent telemetry policies. These mechanisms will be necessary for immersive computing to become socially acceptable at scale. A major open challenge is how to enforce dynamic privacy, ownership, and access-control policies across heterogeneous devices, applications, and administrative domains without compromising usability or real-time performance.

\paragraph{Open testbeds and reproducible ecosystems.}
Long-term progress will depend on open platforms, shared datasets, realistic traces, digital twins, and reproducible benchmarks. Future XR testbeds should support both synthetic and real users, both simulator environments and real devices, real-world 5G/6G deployment with heterogeneous networks, edge/cloud resources, and human-centered evaluation~\cite{du2026vibe,han2023comic}. Such infrastructure and benchmarks~\cite {Popescu2025Benchmarking} will enable the community to move beyond isolated prototypes toward cumulative, rigorous, universally comparable, and application-grounded research.

\paragraph{Key Takeaways.}
\begin{itemize}
    \item Near-term progress in immersive computing will be driven by practical engineering advances in edge--cloud orchestration, adaptive communication, multimodal interaction, and scalable runtime systems.
    
    \item Longer-term evolution may fundamentally reshape networking and computing around semantic communication, persistent shared environments, adaptive infrastructure, and continuously integrated physical--digital interaction.
    
    \item Future immersive systems will increasingly rely on selective intelligence, contextual adaptation, and cross-layer coordination rather than brute-force scaling of bandwidth or computation alone.
    
    \item Realizing these visions requires sustained advances in shared infrastructure, interoperability, trustworthiness, and human-centered system design.
\end{itemize}

\section{Potential Pitfalls and Risks 
}

While the momentum behind AI-empowered immersive computing is strong, the community must navigate several critical risks that could fragment the ecosystem, compromise system resilience, or erode public trust. Workshop discussions highlighted that these pitfalls must be addressed preemptively through cross-disciplinary co-design and open standardization.

\subsection{Over-Reliance on Bandwidth Scaling}

\paragraph{Pitfall.}
A pervasive assumption is that the hurdles of immersive computing will naturally vanish as next-generation networks provide massive bandwidth and ultra-low latency. However, relying solely on network over-provisioning is a fragile strategy. The bottlenecks of immersive systems span sensing, rendering constraints, thermal limits, battery capacity, and human perceptual thresholds. Transmitting denser 3D data often increases energy consumption and privacy exposure without yielding proportional gains in human task performance~\cite{clemm2020toward, strinati20196g}.

\paragraph{Potential solution.}
The community should move from bandwidth-centric design to information-centric and task-aware design. Future frameworks should optimize exactly what data is transmitted, what can be intelligently hallucinated at the edge, and what must remain localized. Leveraging semantic compression, predictive prefetching, and cross-layer scheduling will decouple immersive QoE from the sheer brute force of raw bandwidth scaling~\cite{cheng2024magicstream,zhang2022yuzu,zhang2022m5}.

\subsection{Fragmented Platforms and Lack of Shared Infrastructure}

\paragraph{Pitfall.}
The XR ecosystem remains fragmented across devices, operating systems, APIs, content formats, software stacks, and evaluation methodologies. Consequently, immersive systems research often relies on isolated prototypes and proprietary hardware, making it difficult to reproduce results, compare systems fairly, or evaluate whether proposed techniques generalize to realistic multi-user environments~\cite{han2023comic,Popescu2025Benchmarking}. Researchers and developers must also navigate inconsistent software stacks, privacy models, and performance tradeoffs without common abstractions or shared infrastructure.

\paragraph{Potential solution.}
The community should invest in open, interoperable XR ecosystems that combine shared standards with reproducible infrastructure. This includes common abstractions for computation offloading, sensor access, volumetric content delivery, edge-cloud orchestration, and human-centered evaluation, as well as open-source platforms and standardized benchmarks~\cite{han2023comic,Li2025XR,Popescu2025Benchmarking,schwarz2019standards}. Beyond shared software abstractions, existing networking testbeds should also be extended to support XR workloads, heterogeneous devices, programmable wireless infrastructure, and realistic user traces, enabling scalable and reproducible immersive computing research.

\subsection{Privacy Backlash and Loss of Social Trust}

\paragraph{Pitfall.}
Always-on immersive computing devices are inherently surveillance machines; they continuously capture high-fidelity spatial maps, biometric signatures, eye-gaze telemetry, and audio of both the wearer and passive bystanders. Failure to address the sociotechnical implications of this pervasive sensing will inevitably trigger public backlash, stalling adoption in sensitive domains such as healthcare, education, and public spaces~\cite{Kenwright2018Virtual,shi2025your}.

\paragraph{Potential solution.}
Privacy cannot be a software patch; it must be an architectural primitive. Future platforms require hardware-backed data minimization, localized federated processing, cryptographically secure bystander controls, and highly visible, unspoofable recording indicators. These technical safeguards must be coupled with transparent policy frameworks~\cite{cheng2024metafl,david2024understanding} that enable users to understand and audit how their spatial data is used.

\begin{figure}
    \centering
    \includegraphics[width=1.0\linewidth]{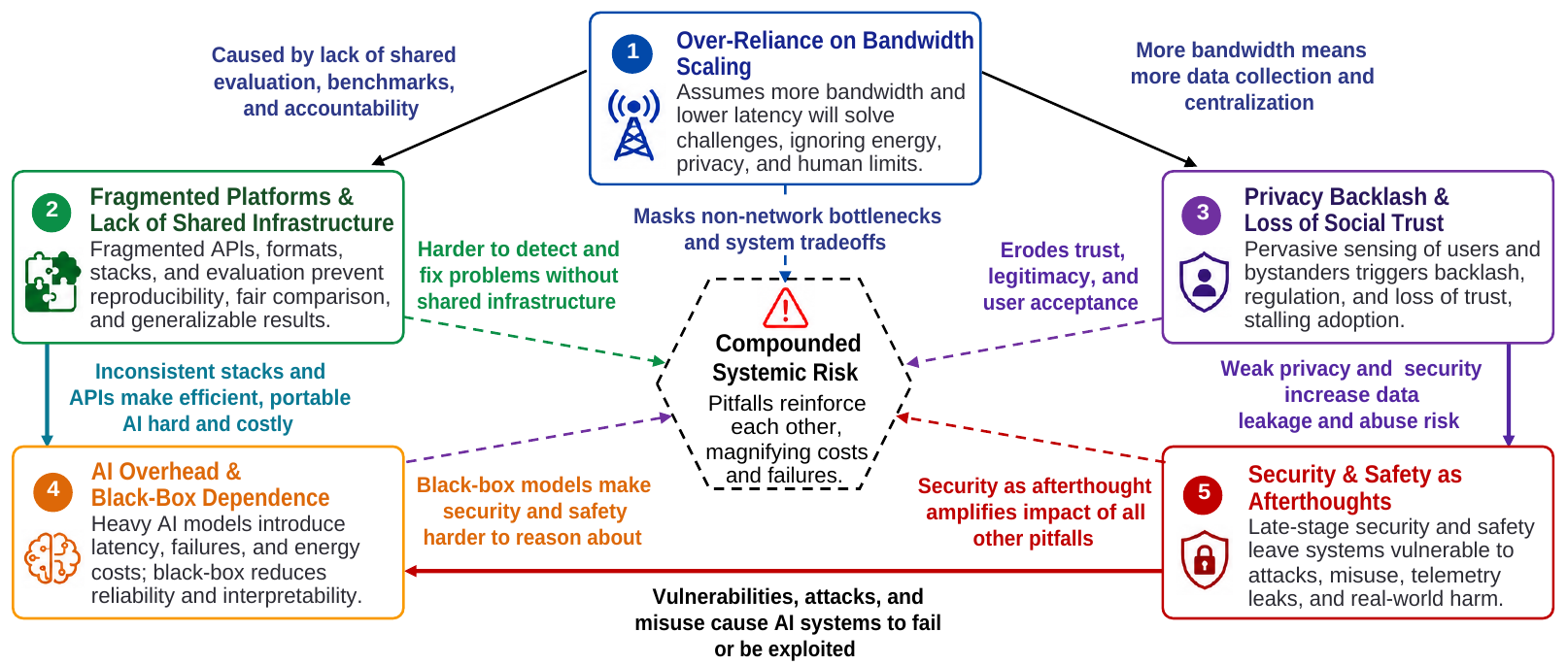}
    \caption{Risk Interdependency Map: How the five key pitfalls amplify one another in immersive computing.
    }
    \label{fig:pitfalls}
\end{figure}

\subsection{AI Overhead and Black-Box Dependence}

\paragraph{Pitfall.}
While deep learning is increasingly positioned as a panacea for immersive computing (e.g., neural rendering, pose estimation, and interaction prediction), deploying foundation models on constrained wearables introduces severe latency, thermal, and energy penalties. Furthermore, neural models are susceptible to catastrophic failures under domain shifts, whereas classical geometric pipelines often offer deterministic, low-latency guarantees~\cite{huynh2017deepmon,lane2016deepx}.

\paragraph{Potential solution.}
AI should be used selectively and co-designed with system constraints. Future XR systems need decision frameworks that balance accuracy, latency, energy, privacy, interpretability, and robustness. Hybrid methods~\cite{cai2021yolobile,fang2018nestdnn} that combine geometry, learning, semantic representations, and application-specific knowledge may be more effective than purely black-box approaches.

\subsection{Security and Safety as Afterthoughts}

\paragraph{Pitfall.}
XR systems can directly influence perception, cognition, and physical behavior. Attacks may manipulate visual, audio, or haptic stimuli; exploit avatars; leak telemetry; induce cybersickness; or affect user safety in physical environments. If security and safety are added only after systems are built, defenses may be ineffective or too costly to deploy.

\paragraph{Potential solution.}
Security and safety should be designed into immersive systems from the beginning. This requires threat models specific to XR, fine-grained access control, secure sensor pipelines, trustworthy rendering, anomaly detection, and evaluation methods that measure degradation, user harm, and safety impact rather than only binary attack success.

\vspace{0.12in}
Taken together, these pitfalls demonstrate that the risks of immersive computing are fundamentally interconnected rather than isolated technical concerns, as shown in Figure~\ref{fig:pitfalls}. Fragmented infrastructure can hinder reproducibility and obscure security weaknesses; bandwidth-centric architectures can amplify privacy exposure and centralized data collection; and black-box AI systems can complicate reliability, safety, and trust. These observations reinforce a central theme of the workshop: immersive computing requires coordinated cross-layer design not only to improve performance and scalability, but also to proactively mitigate systemic risks before they become deeply embedded in future platforms and infrastructure.

\section{What Does Success Look Like in Ten Years? 
}

In ten years, the success of immersive computing should not be measured only by better headsets, higher display resolution, or faster wireless links. A successful research agenda will produce immersive systems that are real-time, scalable, trustworthy, reproducible, human-centered, and useful in everyday settings. The following milestones define a concrete vision for the next decade.

\paragraph{Real-time immersive interaction at scale.}
Immersive systems should support large-scale, real-time interaction among many concurrent users~\cite{liu2024muv2,meng2020coterie,zhang2022m5,zhang2022sear}, including classrooms, training environments, remote collaboration, telepresence, emergency response, and shared digital twins. A successful system should scale beyond isolated single-user experiences toward persistent multi-user environments with synchronized state, shared context, and stable QoE for 100 or more concurrent users.

\paragraph{Sub-50 ms motion-to-photon latency for interactive experiences.}
Future immersive systems should provide consistently low end-to-end latency for motion, rendering, communication, and feedback. The target should not be only low average latency, but stable latency under mobility, wireless variability, changing user behavior, and multi-user load. Achieving sub-50 ms end-to-end latency for interactive scenarios would represent a major step toward natural and comfortable remote presence.

\paragraph{100$\times$ bandwidth reduction through semantic and adaptive communication.}
Success will require moving beyond raw data streaming. Immersive systems should reduce bandwidth demand by transmitting task-relevant information, semantic representations, user intent, scene changes, and prioritized content rather than all pixels or all sensor streams~\cite{cheng2024magicstream}. A 100$\times$ reduction in bandwidth for representative XR workloads would enable broader deployment while reducing energy use and privacy exposure.

\paragraph{Seamless human--AI--environment interaction.}
AI agents should become natural components of immersive environments. They should assist users with instruction, translation, navigation, accessibility, collaboration, and decision-making while remaining transparent, controllable, and privacy-preserving. Success means that users can interact with AI, other humans, and physical or virtual environments in a continuous loop without breaking immersion or increasing cognitive burden.

\paragraph{Widely adopted open testbeds and benchmarks.}
The community should have shared XR testbeds, datasets, traces, benchmarks, and open-source platforms that make results comparable and reproducible. Platforms such as CoMIC~\cite{han2023comic} and other multi-site XR testbeds could support realistic experiments involving real users, heterogeneous devices, edge/cloud resources, programmable networks, and human-centered metrics. These platforms should lower the barrier to entry and allow the field to build cumulative knowledge rather than isolated prototypes.

\paragraph{Privacy-preserving and socially acceptable XR devices.}
In ten years, immersive devices should be deployable in public, professional, educational, and home environments without creating widespread privacy backlash. Users and bystanders should understand when sensing occurs, what data is collected, where it is processed, and how it is protected~\cite{cheng2024metafl,wu2025pipe}. Privacy zones, visible recording indicators, on-device filtering, fine-grained permissions, and policy-aware sensing should become standard platform capabilities.

\paragraph{Application-grounded immersive systems.}
Success also means identifying where XR provides unique value. Mature immersive systems should be validated in compelling domains such as medical training, elder care, education, manufacturing, agriculture, emergency response, and search and rescue. In these domains, XR should improve task performance, training quality, safety, inclusion, or access to expertise compared with conventional alternatives.

\paragraph{Energy-efficient all-day wearable systems.}
For AR glasses and wearable XR devices, success requires all-day usability. Devices should be lightweight, thermally comfortable, socially acceptable, and energy efficient. Computation should be distributed intelligently across glasses, phones, edge devices, and cloud servers so that users receive high-quality immersive experiences without unacceptable battery drain, heat, or device weight.

\paragraph{Standardized interfaces and interoperable ecosystems.}
The field should converge on common abstractions for sensing, rendering, semantic data, offloading, privacy, QoE, and edge/cloud orchestration. These standards should enable interoperability across devices, applications, networks, and platforms while preserving room for innovation. A healthy ecosystem should enable researchers and developers to build portable applications and compare systems on a shared basis.

\paragraph{Trustworthy immersive infrastructure.}
Finally, immersive computing should become part of a trustworthy digital infrastructure. Security, privacy, and safety should be built into the system stack from the beginning. Immersive platforms should protect against perceptual manipulation, telemetry leakage, avatar-based attacks, unsafe interactions, and misuse of sensor data. Trust should be treated not as a single-platform property but as a relationship among users, developers, applications, devices, and physical spaces.

\vspace{0.05in}
Taken together, these milestones describe a future in which immersive computing becomes a practical, responsible, and broadly useful computing paradigm. The goal is not simply to make XR more visually compelling, but to create an immersive Internet that is scalable, adaptive, secure, privacy-preserving, and centered on human needs.

\section{Action Items and Recommendations 
}

Realizing the vision of scalable, trustworthy, and human-centered immersive computing will require coordinated action across academia, industry, government, and the broader research community. The workshop discussions point to a shared conclusion: immersive computing cannot advance through isolated progress in devices, networking, systems, AI, or applications alone. Instead, it requires sustained cross-layer collaboration, shared infrastructure, common evaluation methodologies, and strong connections to real-world deployment needs.

\subsection{Academia}

\paragraph{Advance interdisciplinary, cross-layer immersive systems research.}
Academic researchers should pursue problems that span networking, systems, AI, graphics, HCI, security, privacy, and application domains. Key directions include semantic communication, adaptive edge/cloud orchestration, privacy-aware sensing, hybrid rendering, multi-user synchronization, and human-centered QoE models. Universities should also train interdisciplinary researchers through integrated courses, workshops, and collaborative research programs that bridge traditional disciplinary boundaries.

\paragraph{Develop rigorous and human-centered XR evaluation practices.}
Academic researchers should develop evaluation methodologies that capture both system-level performance and human experience in immersive environments. Beyond conventional metrics such as latency and bandwidth, XR evaluation should incorporate human-centered dimensions, including comfort, accessibility, cognitive load, trust, interaction stability, and task success. Universities can also play an important role in establishing reproducible experimental practices, interdisciplinary evaluation protocols, and open research artifacts that improve the rigor and comparability of immersive computing research.

\paragraph{Integrate domain experts and end users into XR research.}
Immersive systems should be developed in collaboration with stakeholders from domains such as healthcare, education, manufacturing, elder care, and emergency response. Early involvement of end users and domain experts can help identify meaningful deployment opportunities and avoid technology-driven solutions that lack practical value.

\subsection{Industry}

\paragraph{Support interoperable XR standards and interfaces.}
Industry should help develop interoperable standards for XR sensor access, privacy controls, semantic representations, offloading APIs, edge/cloud interfaces, and QoE models. Shared standards are essential for portability across devices and platforms while preserving room for hardware and software innovation.

\paragraph{Share realistic workloads and deployment insights.}
Industrial XR platforms provide valuable insights into real users, applications, devices, and deployment constraints. Companies should share anonymized traces, representative workloads, benchmark applications, and system constraints with the research community to support realistic and reproducible evaluation.

\paragraph{Collaborate on shared infrastructure and open ecosystems.}
Many immersive computing challenges, including interoperability, accessibility, privacy signaling, latency measurement, and multi-user evaluation, are fundamentally pre-competitive. Industry can accelerate progress by partnering with academia and government on shared testbeds, open-source components, and reference platforms.

\subsection{Government and Funding Agencies}

\paragraph{Invest in shared, multi-site XR research infrastructure.}
Funding agencies should support immersive computing infrastructure that integrates XR devices, programmable wireless networks, edge/cloud resources, realistic workloads, and human-subject evaluation. These platforms should enable large-scale, multi-user, and geographically distributed experiments that cannot be reproduced in isolated laboratory environments.

\paragraph{Support interdisciplinary immersive computing programs.}
Immersive computing spans networking, systems, AI, graphics, HCI, security, privacy, cognitive science, and domain applications. Funding programs should encourage cross-disciplinary teams and support projects that combine technical innovation with human-centered evaluation and real-world deployment.

\paragraph{Embed trustworthy computing into core XR research.}
Privacy, security, safety, accessibility, and societal impact should be integrated into mainstream immersive computing programs rather than treated as secondary concerns. Funding agencies can play a critical role by prioritizing trustworthy and human-centered XR research from the outset.

\subsection{Community}

\paragraph{Make reproducibility and artifact sharing standard practice.}
Immersive computing publications should increasingly include code, datasets, traces, configuration files, user-study protocols, and experimental artifacts. Establishing reproducibility as a community norm will improve scientific rigor and enable more reliable comparisons across systems and methodologies.

\paragraph{Converge on common terminology and evaluation metrics.}
The community should establish shared definitions for latency, QoE, immersion, interaction stability, privacy risk, semantic fidelity, and multi-user consistency. Common terminology and metrics are essential for communication across disciplines and for meaningful evaluation of immersive systems.

\paragraph{Sustain long-term cross-community collaboration.}
Recurring workshops, benchmark competitions, interdisciplinary meetings, and community roadmaps can help align researchers across networking, systems, AI, XR, graphics, HCI, security, and application domains. These efforts should prioritize shared infrastructure, open evaluation, and long-term research coordination.

\section{Conclusion 
}
The future of computing may be shaped less by screens and applications and more by persistent, shared, and intelligent digital experiences embedded into the physical world. Immersive computing is beginning to move beyond isolated XR demonstrations toward a new paradigm in which people, AI agents, devices, and environments interact continuously through spatial, multimodal interfaces. Enabling this transition will require far more than incremental advances in displays or wireless bandwidth. It demands a fundamental redesign of the networking, systems, and infrastructure that support real-time, distributed, sensor-rich, and human-centered interaction at scale.

Unlike traditional applications, immersive experiences tightly couple sensing, AI, compute, networking, rendering, and human interaction into a single closed-loop system. As a result, immersive computing cannot be built through isolated optimizations at individual layers of the stack. Decisions about representation, inference, compression, offloading, or synchronization immediately affect latency, scalability, energy consumption, privacy, and user experience. For example, semantic scene representations may dramatically reduce communication overhead while increasing computational complexity~\cite{cheng2024magicstream}; edge-assisted rendering may improve device efficiency but introduce new challenges for trust and responsiveness~\cite{wu2024theia}; generative reconstruction may compensate for network loss while raising concerns about transparency and reliability~\cite{wu2025nevo}. The central challenge, therefore, is no longer optimizing individual components, but co-designing end-to-end immersive systems that jointly balance performance, scalability, efficiency, trustworthiness, and QoE~\cite{han2026ai}.

Addressing these challenges will require sustained coordination across academia, industry, government, and the broader research community. Academic researchers must establish the theoretical foundations, architectures, benchmarks, and evaluation methodologies for immersive computing, while training the next-generation workforce to work across traditionally separated areas such as networking, systems, AI, graphics, HCI, and security. Industry will play a critical role in defining interoperable platforms, deployment practices, device interfaces, and privacy-preserving APIs that allow immersive applications to scale beyond vertically integrated ecosystems. Government agencies can accelerate progress by supporting shared research infrastructure, open datasets, interdisciplinary collaborations, large-scale experimentation, and long-term community building. Equally important, the broader community must develop common abstractions, reproducible methodologies, and open platforms that enable cumulative scientific progress rather than fragmented one-off prototypes.

Ultimately, immersive computing should be viewed not as a standalone technology domain, but as a new societal computing substrate. Future immersive systems will operate continuously across homes, hospitals, classrooms, factories, workplaces, and public spaces, mediating how people communicate, learn, collaborate, and interact with AI. These systems will sense both users and their surrounding environments, making decisions that directly influence perception, attention, and behavior. Consequently, privacy, security, accessibility, fairness, social acceptability, and human-centered evaluation cannot remain secondary considerations or afterthoughts. They must become foundational design principles embedded into immersive systems from the beginning.

Achieving this vision will require coordinated efforts across the following areas.

\begin{itemize}[leftmargin=*]

\item Establish dedicated cross-layer immersive computing research programs that treat networking, systems, AI, XR, HCI, graphics, security, and privacy as tightly coupled components of a single end-to-end stack.

\item Fund and sustain open XR testbeds that support real users, heterogeneous devices, programmable networks, edge/cloud resources, realistic workloads, multi-user scenarios, and human-centered evaluation.

\item Create shared datasets, traces, benchmarks, and digital twins for immersive systems, including sensor traces, network traces, user-motion traces, volumetric content, interaction logs, and application-specific scenarios.

\item Define common metrics for immersive QoE that go beyond bitrate, frame rate, and visual fidelity to include latency stability, cybersickness, cognitive load, comfort, accessibility, task success, privacy expectations, trust, and social acceptability.

\item Promote semantic and adaptive communication techniques that reduce unnecessary data movement by transmitting task-relevant information, scene structure, user intent, object semantics, or prioritized content rather than raw pixels alone.

\item Orchestrate real-time edge--cloud frameworks that dynamically decide what to compute, where to compute it, what to transmit, what to cache, and what to render based on latency, power, privacy, reliability, and application needs.

\item Build privacy-preserving and trustworthy XR platforms with data minimization, on-device processing, fine-grained sensor permissions, visible recording indicators, secure telemetry handling, and mechanisms for bystander awareness and control.

\item Support human-centered and longitudinal in-situ studies that involve domain experts, end users, and affected communities early in the design process, especially in healthcare, education, assistive living, emergency response, and public-space applications.

\item Encourage open-source XR systems, reference implementations, shared evaluation pipelines, and artifact submission practices that make immersive computing research reproducible and comparable across institutions.

\item Maintain a living community roadmap for immersive computing that tracks emerging applications, open challenges, infrastructure needs, interoperability standards, evaluation methods, and long-term research milestones.

\end{itemize}

If these efforts succeed, immersive computing can become more than a set of compelling prototypes. It can become a trustworthy, scalable, reproducible, and human-centered infrastructure for the next generation of digital interaction. 
The opportunity is significant: {\em to build an immersive Internet that allows people, AI agents, and physical environments to interact seamlessly, safely, and meaningfully}. 
Achieving this vision will require treating immersive computing not as an application at the edge of today's Internet, but as a driving force for rethinking the Internet and computing systems of tomorrow.

\section{Acknowledgment}
This report emerged from an NSF-sponsored workshop titled ``{\em The NSF Workshop on Networking and Systems Challenges in Immersive Computing}''. 
In addition to the authors, participants and contributors of this workshop include 
Isaac Amundson,
Aryabrata Basu,
Jacob Chakareski,
Yasra Chandio,
Kavitha Chandra,
Bo Chen,
Yingying (Jennifer) Chen,
Mallesham Dasari,
Brendan David-John,
Saptarshi Debroy,
Ashutosh Dhekne,
Sonia Fahmy,
Maria Gorlatova,
Tian Guo,
Qi Han,
Bo Ji,
Jaehong Kim,
Dmitrii Korobeinikov,
Bin Li,
Ming Li,
Alan Zaoxing Liu,
Yao Liu,
Yu Liu,
Hurriyet Ok,
Eman Ramadan,
Leon Reznik,
Paul Ruth,
Srinivasan Seshan,
Vishrant Tripathi,
Wei Wang,
Sheng Wei,
Peng Wu,
Zhisheng Yan,
Raman Zatsarenko,
Hongwei Zhang,
Lixia Zhang,
Xiaokuan Zhang,
Zhuangdi Zhu.
We acknowledge NSF and the following NSF program directors, Jason O. Hallstrom, Hang Liu, Claire Lunde, Abraham Matta, Mimi McClure, Deepankar (Deep) Medhi, and Greg Peterson, for their support of this workshop. 
The presentation slides, video recordings, and summaries of the breakout discussion sessions at the workshop are available on the following website: \url{https://immercom.github.io/nsf/immersive}.

\printbibliography

\end{document}